\documentclass[lettersize,journal]{IEEEtran}
\usepackage{amsmath,amsfonts}
\usepackage{algorithmic}
\usepackage{algorithm}
\usepackage{array}
\usepackage[caption=false,font=normalsize,labelfont=sf,textfont=sf]{subfig}
\usepackage{textcomp}
\usepackage{stfloats}
\usepackage{url}
\usepackage{verbatim}
\usepackage{graphicx}
\usepackage{cite}
\usepackage{orcidlink}
\usepackage{textgreek}
\usepackage{amsmath, bm, mathabx}
\usepackage{braket}

\begin{document}

\title{Design and simulation of a transmon qubit chip for Axion detection}
\author{R.~Moretti \orcidlink{0000-0002-5201-5920},
        H.~A.~Corti \orcidlink{0000-0002-5335-0042},
        D.~Labranca \orcidlink{0000-0002-5351-0034},
        F.~Ahrens \orcidlink{0009-0003-6609-795X},
        G.~Avallone \orcidlink{0000-0001-5482-0299},
        D.~Babusci \orcidlink{0000-0003-1312-3272},
        L.~Banchi \orcidlink{0000-0002-6324-8754},
        C.~Barone \orcidlink{0000-0002-6556-7556},
        M.~M.~Beretta \orcidlink{0000-0002-7026-8171},
        M.~Borghesi \orcidlink{0000-0001-5854-8894},
        B.~Buonomo \orcidlink{0000-0002-3612-7308},
        E.~Calore \orcidlink{0000-0002-2301-3838},
        G.~Carapella \orcidlink{0000-0002-0095-1434},
        F.~Chiarello \orcidlink{0000-0002-7487-2827},
        A.~Cian \orcidlink{0000-0003-3497-7066},
        A.~Cidronali \orcidlink{0000-0002-1064-7305},
        F.~Costa \orcidlink{0000-0001-7572-2014},
        A.~Cuccoli \orcidlink{0000-0002-2556-9944},
        A.~D'Elia \orcidlink{0000-0002-6856-7703},
        D.~Di Gioacchino \orcidlink{0000-0002-1288-4742},
        S.~Di Pascoli \orcidlink{0000-0001-6556-7815},
        P.~Falferi \orcidlink{0000-0002-1929-4710},
        M.~Fanciulli \orcidlink{0000-0003-2951-0859},
        M.~Faverzani \orcidlink{0000-0001-8119-2953},
        G.~Felici \orcidlink{0000-0001-8783-6115},
        E.~Ferri \orcidlink{0000-0003-1425-3669},
        G.~Filatrella \orcidlink{0000-0003-3546-8618},
        L.~G.~Foggetta \orcidlink{0000-0002-6389-1280},
        C.~Gatti \orcidlink{0000-0003-3676-1787},
        A.~Giachero \orcidlink{0000-0003-0493-695X},
        F.~Giazotto \orcidlink{0000-0002-1571-137X},
        D.~Giubertoni \orcidlink{0000-0001-8197-8729},
        V.~Granata \orcidlink{0000-0003-2246-6963},
        C.~Guarcello \orcidlink{0000-0002-3683-2509},
        G.~Lamanna \orcidlink{0000-0001-7452-8498},
        C.~Ligi \orcidlink{0000-0001-7943-7704},
        G.~Maccarrone \orcidlink{0000-0002-7234-9522},
        M.~Macucci \orcidlink{0000-0002-7943-2441},
        G.~Manara \orcidlink{0000-0002-1242-1908},
        F.~Mantegazzini \orcidlink{0000-0002-5620-2897},
        P.~Marconcini \orcidlink{0000-0003-3714-0026},
        B.~Margesin \orcidlink{0000-0002-1120-3968},
        F.~Mattioli \orcidlink{0000-0002-7242-3366},
        A.~Miola \orcidlink{0000-0003-0740-5070},
        A.~Nucciotti \orcidlink{0000-0002-8458-1556}, 
        L.~Origo \orcidlink{0000-0002-6342-1430},
        S.~Pagano \orcidlink{0000-0001-6894-791X},
        F.~Paolucci \orcidlink{0000-0001-8354-4975},
        L.~Piersanti \orcidlink{0000-0003-3186-3514},
        A.~Rettaroli \orcidlink{0000-0001-6080-8843},
        S.~Sanguinetti \orcidlink{0000-0002-4025-2080},
        S.~F.~Schifano \orcidlink{0000-0002-0132-9196},
        P.~Spagnolo \orcidlink{0000-0001-7962-5203},
        S.~Tocci \orcidlink{0000-0002-5800-5408},
        A.~Toncelli \orcidlink{0000-0003-4400-8808},
        G.~Torrioli \orcidlink{0000-0002-2298-304X},
        A.~Vinante \orcidlink{0000-0002-9385-2127}
\thanks{R.~Moretti, D.~Labranca, M.~Borghesi, M.~Faverzani, A.~Giachero, A.~Nucciotti, L.~Origo
        are with 
        Dipartimento di Fisica, Universit\`{a} di Milano-Bicocca, also with
        INFN - Sezione di Milano Bicocca, I-20126, Milano, Italy, and also with
        Bicocca Quantum Technologies (BiQuTe) Centre, I-20126, Milano, Italy. 
        (Corresponding author: R.~Moretti.)
        }
\thanks{ M.~Fanciulli and S.~Sanguinetti  
        are with 
        Dipartimento di Scienza di Materiali, Universit\`{a} di Milano-Bicocca, also with
        INFN - Sezione di Milano Bicocca, I-20126, Milano, Italy, and also with
        Bicocca Quantum Technologies (BiQuTe) Centre, I-20126, Milano, Italy. 
        }
\thanks{F. Ahrens, A. Cian, D. Giubertoni, F. Mantegazzini, and B. Margesin are with
        Fondazione Bruno Kessler (FBK), I-38123, Trento, Italy, also with
        INFN - TIFPA, I-38123, Trento, Italy.
        }
\thanks{P. Falferi and A. Vinante are with Fondazione Bruno Kessler, also with CNR IFN,
and also with INFN - TIFPA, I-38123, Povo, Trento, Italy.}
\thanks{E.~Ferri is with
        INFN - Sezione di Milano Bicocca, I-20126, Milano, Italy.
        }
\thanks{F.~Chiarello, F.~Mattioli, G.~Torrioli are with
        Istituto di Fotonica e Nanotecnologie - CNR, I-00133 Roma, Italy, and also with INFN - Laboratori Nazionali di Frascati, I-00044, Frascati, Roma, Italy.
        }
\thanks{D.~Babusci, M.~M.~Beretta, B.~Buonomo, A.~D'Elia, D.~Di Gioacchino, G.~Felici, L.~G.~Foggetta, C.~Gatti, C.~Ligi, G.~Maccarrone, L.~Piersanti, A.~Rettaroli, S.~Tocci are with
        INFN - Laboratori Nazionali di Frascati, I-00044, Frascati, Roma, Italy.
        }
\thanks{H.~A.~Corti is with Dipartimento di Ingegneria dell'Informazione, Università di Pisa, I-56122 Pisa, Italy, also with INFN - sezione di Firenze, I-50019, Sesto Fiorentino, Firenze, Italy.}
\thanks{L.~Banchi and A.~Cuccoli are with
				Dipartimento di Fisica e Astronomia, Università di Firenze, and also with 
				INFN Sezione di Firenze, 
                    I-50019, Sesto Fiorentino, Firenze, Italy. 
        }
\thanks{A.~Cidronali is with Dipartimento di Ingegneria dell'Informazione, Università di Firenze, and also with INFN - Sezione di Firenze, I-50019, Sesto Fiorentino, Firenze, Italy.
        }
\thanks{A.~Miola, S.~F.~Schifano are with 
        Università degli Studi di Ferrara, and also with INFN - Sezione di Ferrara,  I-44122, Ferrara, Italy.
        }
\thanks{E.~Calore is with INFN - Sezione di Ferrara, I-44122, Ferrara, Italy.
        }
\thanks{F.~Costa, S.~Di Pascoli, M.~Macucci, G.~Manara, P.~Marconcini are with Dipartimento di Ingegneria dell’Informazione, Università
di Pisa, I-56122, Pisa, Italy.
        }
\thanks{F.~Giazotto is with NEST Istituto Nanoscienze-CNR and also with Scuola Normale Superiore, I-56127, Pisa, Italy.
        }
\thanks{G.~Lamanna, A.~Toncelli are with Dipartimento di Fisica, Università di Pisa, also with INFN - Sezione di Pisa, I-56127, Pisa, Italy.
        }
\thanks{F.~Paolucci, P.~Spagnolo are with INFN - sezione di Pisa, I-56127, Pisa, Italy.
        }
\thanks{G.~Avallone, C.~Barone,  G.~Carapella, V.~Granata, C.~Guarcello, S.~Pagano are with Dipartimento di Fisica, Università di Salerno, also with INFN - sezione di Salerno, I-84084, Fisciano, Salerno, Italy.
        }
\thanks{G.~Filatrella is with Dipartimento di Scienze e Tecnologie, Università del Sannio, I-82100, Benevento, Italy, also with INFN - sezione di Salerno, I-84084, Fisciano, Salerno, Italy.}
}

\markboth{MANUSCRIPT TRACKING ID: EUCAS23-1-EO-DB-03S}
{Shell \MakeLowercase{\textit{et al.}}: A Sample Article Using IEEEtran.cls for IEEE Journals}

\maketitle
This work is licensed to
IEEE under the Creative Commons Attribution 4.0(CCBY 4.0).\\
© © 2024 IEEE. Personal use of this material is permitted. Permission from IEEE must be obtained for all other uses, in any current or future media, including reprinting/republishing this material for advertising or promotional purposes, creating new collective works, for resale or redistribution to servers or lists, or reuse of any copyrighted component of this work in other works.

\begin{abstract}
Quantum Sensing is a rapidly expanding research field that finds one of its applications in Fundamental Physics, as the search for Dark Matter. Devices based on superconducting qubits have already been successfully applied in detecting few-GHz single photons via Quantum Non-Demolition measurement (QND). This technique allows us to perform repeatable measurements, bringing remarkable sensitivity improvements and dark count rate suppression in experiments based on high-precision microwave photon detection, such as for Axions and Dark Photons search. In this context, the INFN Qub-IT project goal is to realize an itinerant single-photon counter based on superconducting qubits that will exploit QND for enhancing Axion search experiments. 
In this study, we present Qub-IT's status towards the realization of its first superconducting qubit device, illustrating design and simulation procedures and the characterization of fabricated Coplanar Waveguide Resonators (CPWs) for readout. We match target qubit parameters and assess a few-percent level agreement between lumped and distributed element simulation models. We reach a maximum internal quality factor of 9.2\texttimes 10\textsuperscript{5} for -92 dBm on-chip readout power.
\end{abstract}

\begin{IEEEkeywords}
Qubit design, Quantum sensing, Qubit simulation, CPW characterization, X-mon
\end{IEEEkeywords}

\section{Introduction}
Recent advancements in the field of quantum physics have enabled significant progress in the precise measurement and control of individual quanta such as microwave-photons \cite{schuster}, phonons \cite{o’connel____2010,2015,Qacustics,Qbar,qubitbaw, remoteentangl} and magnons \cite{magn1, magn2}, opening new
directions in the detection of Dark Matter and Fifth Forces \cite{dm1,dm2,dm3}, in tests of
Quantum Gravity \cite{qg1,qg2} and Quantum Mechanics of macroscopic objects \cite{macro1,macro2}.

One remarkable feature of quantum systems is their capability to preserve the history of their interactions with the environment within their quantum state. This property can find a practical application in the form of Quantum Non-Demolition (QND) detection, which proves to be well-suited for implementation in superconducting qubits platforms for single microwave photon detection \cite{dm1,qnd1}. This is exemplified by considering a superconducting qubit, initially prepared in a superposition of ground $\ket{g}$ and excited $\ket{e}$ state:
\begin{equation}
\ket{\psi} = \frac{1}{\sqrt{2}} \left( \ket{g} + \ket{e} \right)
\end{equation} 
In addition to a dispersive readout resonator coupling for readout, the qubit is also weakly coupled, to a high-quality factor cavity which serves the purpose of photon storage. Using this experimental setup, when resonant photons enter the storage cavity, the qubit undergoes a time-dependent phase shift represented by the state:
\begin{equation}
\ket{\psi'} = \frac{1}{\sqrt{2}} \left( \ket{g} + e^{-i2n\xi t}\ket{e} \right)
\end{equation}
Here, $\xi$ is determined by the qubit-cavity coupling strength, and $n$ corresponds to the number of photons stored in the cavity. By conducting Ramsey interferometry on the qubit, we can infer the presence of photons in the cavity, without destroying the photon state. Thanks to the QND nature of this technique, redundancy in photon measurements makes it possible to suppress false positives in single-photon counting experiments exponentially.


The Qub-IT project aims to advance quantum sensing using superconducting qubits for fundamental physics, with a primary focus on axion search, hence its direct application to experiments such as QUAX \cite{quax}.
The axion-photon coupling mechanism only occurs in the presence of magnetic fields, which presents an added complexity for the measurement setup. To address this challenge, our approach will involve itinerant-photon detection, performing QND measurements on the photon as it travels along a transmission line, from the region with the intense magnetic field to the qubit system one \cite{itinerant1,itinerant2}. Magnetic field suppression around the qubit chip region is achievable through spatial separation between the storage cavity and the photon counter subsystems, linked by a coherence-preserving waveguide. A similar technique was demonstrated in \cite{magnard} for achieving remote entanglement between qubits stored in different refrigerators.

In this contribution, we present the status of Qub-IT \cite{Labranca_2023}, towards the fabrication of its first transmon \cite{transmon} qubit chip. Sec. \ref{sec:calculations} presents an analytical model for estimating target Hamiltonian parameters. In Sec. \ref{sec:design} we showcase the final design and the simulation technique to extract all the relevant parameters needed for control and readout. Sec. \ref{sec:cpw} highlights the current advancements in fabricating the readout coplanar waveguide resonators (CPW). We draw our conclusions in Sec. \ref{sec:conclusion} and 

\section{Analytic model}\label{sec:calculations}
The interaction Hamiltonian can be constructed following the approach of 
Ref.~\cite{rasmussen2021superconducting}. Starting from the circuit, e.g. that
presented in Fig.~\ref{fig:lumped},
we first introduce a classical Lagrangian 
\begin{equation}
	\mathcal L= \frac {\dot{\vec \Phi}\cdot C \dot{\vec \Phi}}2 - \frac{ \vec \Phi \cdot L^{-1} \vec \Phi}2 + 
	E_j \cos\left(\frac{2\pi}{\Phi_0}{\phi}\right),
\end{equation}
where $\vec\Phi=(\phi_{\rm ext}, \phi_r,\phi)$ is the vector of fluxes of the external drive, 
resonator and qubit, respectively; $E_j$ is the qubit Josephson energy; $C$ is the capacitance 
matrix, whose entries depend on the capacitive coupling between different nodes, and $L$ is the inductance 
matrix. Following standard quantization procedures, we obtain the quantum Hamiltonian 
\begin{align}
	\hat H  =& \frac{\hat Q^2}{2 \tilde C_t} - 	E_j \cos\left(\frac{2\pi}{\Phi_0}{\hat \phi}\right) 
	+ \frac{\hat Q_r^2}{2 \tilde C_r} + \frac{\hat \phi_r^2}{2 L_r}  + 
	\label{eq:qubit resonator}
	 \\ &+\frac{\hat Q \hat Q_r}{\tilde C_{rt}}
 + (\tilde \beta_t \hat Q + \tilde \beta_r \hat Q_r)V(t),
 \label{eq:full driving}
\end{align}
where $\hat Q$ are the charge operators, the canonical conjugate of the flux.  
The terms in \eqref{eq:qubit resonator} respectively describe the qubit and the
resonator, while those in \eqref{eq:full driving} 
describe the interaction between them and with the external drive. Explicit expressions 
for the coefficients as a function of the circuit elements will be presented elsewhere. 
From those coefficients, we can compute the qubit charge energy as $E_C=e^2/(2\tilde C_t)$. 

The qubit Hamiltonian can be diagonalized as $\sum_n E_n\ket n\!\bra n$ and in the transmon regime we can keep only 
the lowest energy levels, which will be denoted as $\ket0$ and $\ket1$. Their 
difference in energy is denoted as $\hbar\omega_q$, while the difference between $E_2-E_1$ and 
$\hbar \omega_q$ defines the qubit anharmonicity $\alpha$. The resonator Hamiltonian 
can be written as $\hbar \omega_r\hat a^\dagger \hat a$, with resonator frequency $\omega_r$. 
When the qubit and the resonator are off-resonance, after a rotating wave approximation their 
interaction can be written  as 
\begin{equation}
	\frac{\hat Q \hat Q_r}{\tilde C_{rt}}   
	\simeq \hbar g_{\rm tr} \left(\hat a^\dagger \ket0\!\bra1 + \hat a \ket1\!\bra0\right),
\end{equation}
with coupling $g_{\rm tr}$. Moreover, in the dispersive limit, with  a Schrieffer-Wolff 
transformation we can decouple the qubit and resonator to get the following 
effective interaction 
\begin{equation}
	\hat H \simeq -\frac{\hbar \tilde \omega_{q}}2\hat \sigma^z + \hbar  (\tilde \omega_r-\chi \hat \sigma^z)\hat a^\dagger \hat a,
\end{equation}
where the new frequencies have a correction due to the coupling $g_{\rm tr}$ and $\chi$ is 
called dispersive shift. 

\section{Design prototype and simulation}\label{sec:design}
\subsection{Design}
\begin{figure}
    \centering
    \includegraphics[width=0.4\textwidth]{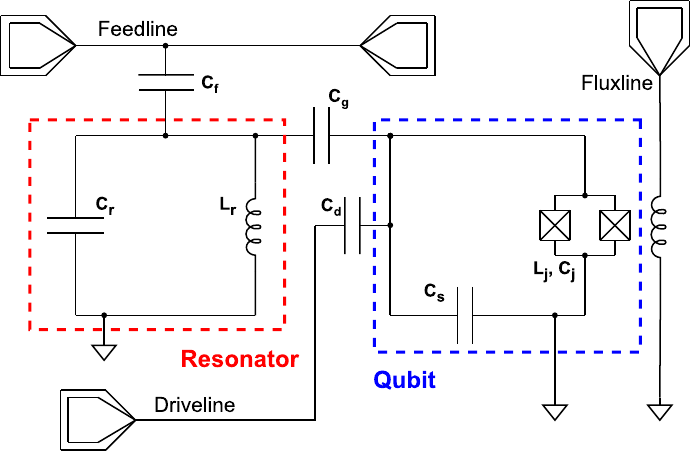}
    \caption{Lumped-element representation of the full design circuit. The resonator is modelled as an LC circuit while the qubit is a SQUID shunted with a capacitance.}
    \label{fig:lumped}
\end{figure}

We present the design details of a planar transmon qubit chip, which is coupled to a transmission line (feedline) through a $\lambda / 4$ resonator. We use aluminium as the superconductor on top of a $600\,$\textmu m silicon substrate. The design comprises a base version and three improved variations. The base version consists only of a fixed-frequency, resonator-driven transmon qubit. In the first variation, we introduce a dedicated driveline to enable faster qubit control, while the second variation features a flux-bias line, which allows tuning of the energy spacing between the qubit excitation levels. Lastly, the third version includes both the driveline and the flux-bias line, combining the advantages of both setups. The lumped-element circuit is shown in Fig.~\ref{fig:lumped}. All four versions share the same qubit-resonator-feedline geometry.
The qubit consists of a nonlinear inductor (a Josephson junction for the fixed-frequency transmon, a DC-SQUID in the flux-tunable transmon case) connected in parallel with a shunt capacitor, formed by a cross-shaped pad and the ground plane. Such a configuration is also called Xmon \cite{xmon}. The shunt capacitor allows for charge noise mitigation \cite{transmon} and must be carefully tuned in order to reach the transmon regime. The CPW resonator is capacitively coupled to the feedline and one of the qubit crosses. The latter is mediated by a capacitive claw that customises the coupling strength by tuning the claw length and distance from the shunt capacitor.

\begin{figure}
    \centering
    \includegraphics[width = 0.38\textwidth]{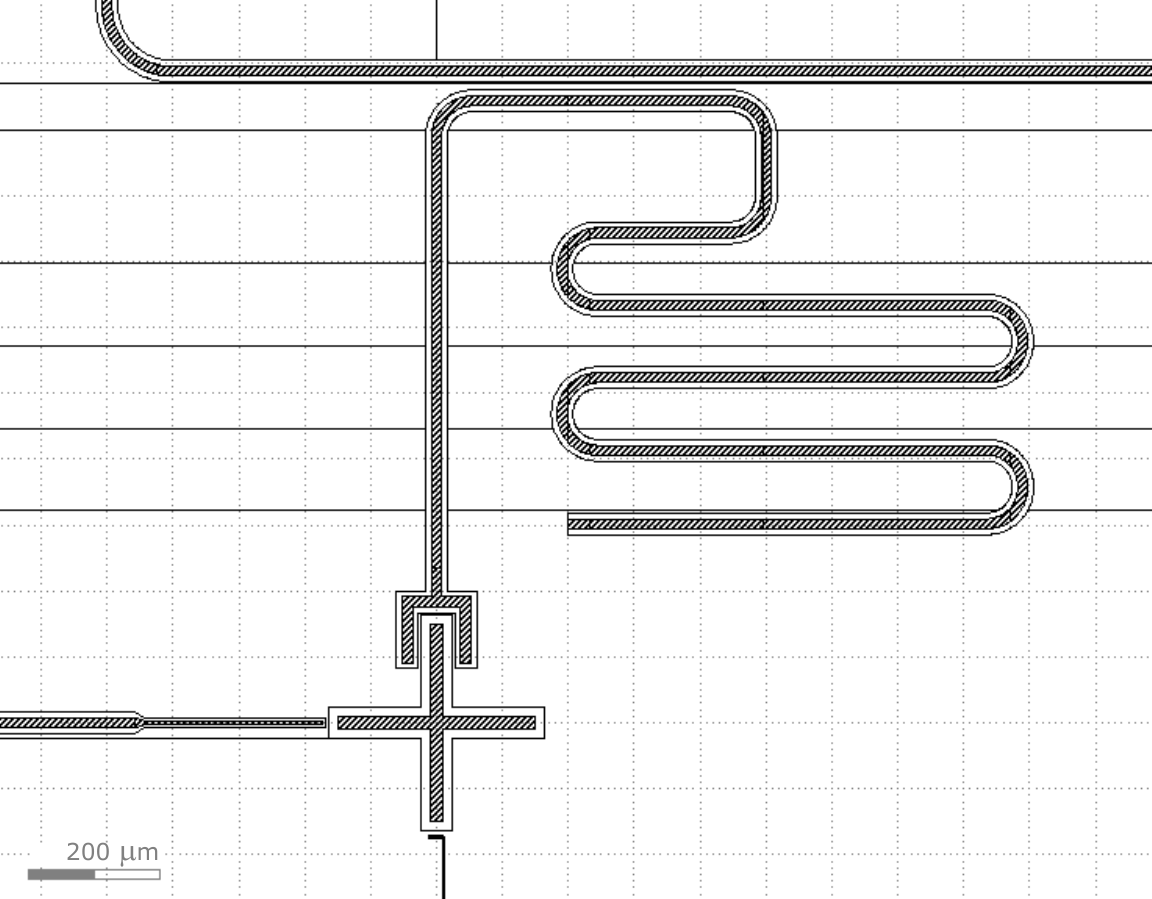}
    \caption{Close-up view of the chip rendering. The Xmon couples with the CPW resonator claw, the driveline and the flux-bias line, respectively through the top, left and bottom arms. The resonator is connected to ground and additionally couples to the feedline.}
    \label{fig:design_gds}
\end{figure}

We developed the circuit design using the Qiskit-Metal toolkit \cite{qiskitmetal}, and it is depicted in Fig.~\ref{fig:design_gds}. 
The Xmon cross has dimensions $21\,$\textmu m width and $342\,$\textmu m length, with a gap of $14\,$\textmu m from the ground plane. The CPW readout resonator measures $4.689\,$mm in length, with a trace width of $15\,$\textmu m and a $9\,$\textmu m gap, ensuring a characteristic impedance of $Z_0 = 50\,$\textOmega. The coupling element is an $85\,$\textmu m long claw with the same trace width and gap of the resonator. Additionally, the resonator is capacitively coupled to the feedline through a $500\,$\textmu m coupling trait $30\,$\textmu m distant from the feedline.

\subsection{Simulation}
The simulations were executed using Ansys Q3D and Ansys HFSS software. Ansys Q3D was utilized for extracting capacitance values tied to distinct circuit elements and couplings, which are summarized in Table \ref{tab:cs}, assuming a parasitic capacitance for the non-linear inductor equal to $c_\text{J}=2\,$fF.

\begingroup
\setlength{\tabcolsep}{8pt} 
\renewcommand{\arraystretch}{1.2} 
\begin{table}[htbp]
    \caption{Capacitance values extracted with Ansys Q3D.}
    \centering
    \begin{tabular}{|c|c|c|c|c|}
    \hline
         $\mathbf{c_s}$ & $\mathbf{c_g}$ & $\mathbf{c_f}$ & $\mathbf{c_d}$ & $\mathbf{c_r}$\\
         \hline 
         $87.40\,$fF& $3.93\,$fF& $6.41\,$fF& $0.20\,$fF& $404.07\,$fF\\ \hline
    \end{tabular}
    \label{tab:cs}
\end{table}
\endgroup
Further use of Q3D was the extraction of the mutual inductance between the SQUID and the flux-bias line, resulting in $2.7\,$mA per induced flux quantum.

For quantization with Q3D-extracted data, we resorted to the Lumped Oscillator Model (LOM) \cite{lom} analysis. The latter was carried out using the capacitance matrix related to the qubit, resonator claw and ground plane subsystem. For taking into account dielectric losses and $50\,$\textOmega\ port impedances,
as well as to obtain more accurate estimates of resonator and qubit frequencies with their quality factors, we resorted to Ansys HFSS performing eigenmode analysis. Quantization in this case consisted of the Energy Participation Ratio (EPR) method \cite{epr}, which quantifies how much of the energy of a mode is stored in each element to evaluate several parameters of interest, such as the Kerr coefficients:

\begin{equation}
    \chi_{nm} = \frac{\hbar\omega_m\omega_n}{4E_{\text{J}}}p_m p_n
\end{equation}
where the anharmonicities are $\alpha_m = \chi_{mm}/2$ and the total dispersive shifts $\chi_{nm}$ for $n\neq m$. Here, $p_m$ and $p_n$ refer to the energy participation ratios of the Josephson element for modes $m$ and $n$. For simplicity, we will refer to $\chi$ as the resonator total dispersive shift induced by qubit transitions $\ket{g} \leftrightarrow \ket{e}$, and to $\alpha$ the qubit's anharmonicity.

In addition, with the eigenmode simulation, we evaluated the contributions of the driveline and flux-bias line, verifying that they do not affect significantly the Kerr coefficients of the cavity and the qubit.

Assuming a zero-bias Josephson inductance of $L_\text{j}=10\,$nH, matching a critical current $I_c=33.0\,$nA, the results obtained with the LOM and the EPR methods agree to the amount expected \cite{lom,epr} and are reported in table \ref{tab:results}. The LOM analysis needs the dressed resonator frequency
$\omega_r$ as an input, therefore we employed the value extracted through the EPR method.

\begingroup
\setlength{\tabcolsep}{8pt} 
\renewcommand{\arraystretch}{1.2} 
\begin{table}[htpb]
    \caption{Results extracted from LOM and EPR simulations, assuming $L_\text{j}=10$ nH and $C_\text{j}=2$ fF.}
    \centering
    \begin{tabular}{|c|c|c|}
    \hline 
         & \textbf{LOM} & \textbf{EPR}\\ \hline 
        Transmon regime $E_\text{j} / E_\text{c}$ & $79.44$ & $80.92$ \\ \hline 
        Anharmonicity $\alpha/2\pi$ & $227.81\,$MHz & $213.80\,$MHz\\ \hline
        Dispersive shift $\chi/2\pi$ & $0.558\,$MHz & $0.438\,$MHz\\ \hline
        Qubit frequency $\omega_q/2\pi$ & $4.970\,$GHz & $4.878\,$GHz \\ \hline
        Cavity frequency $\omega_r/2\pi$ & $5.989\,$GHz & $5.988\,$GHz\\ \hline
    \end{tabular}

    \label{tab:results}
\end{table}
\endgroup

\begin{figure}
    \centering
    \includegraphics[width = 0.45\textwidth]{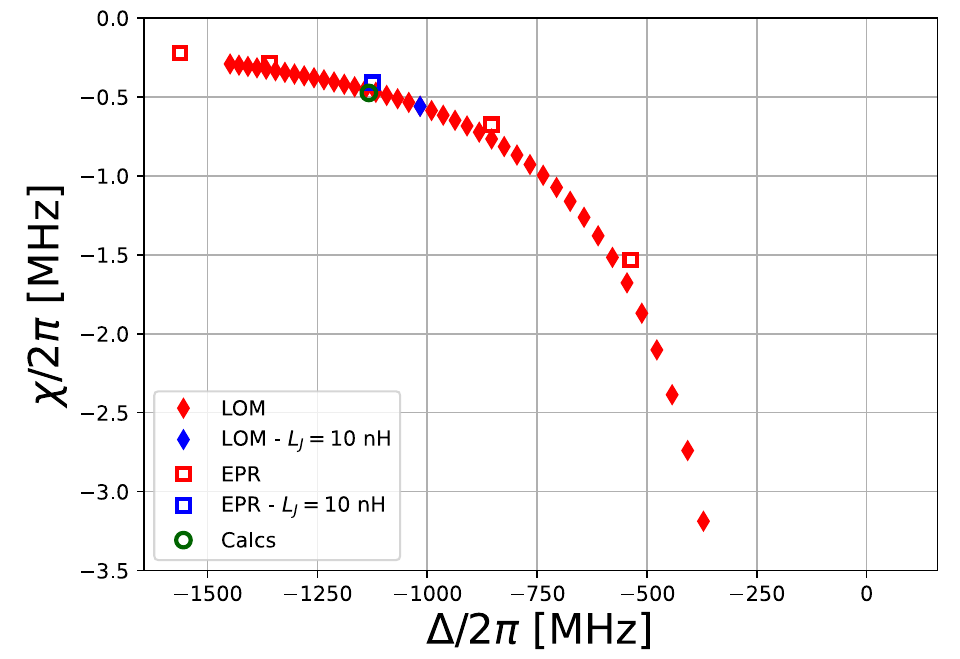}
    \caption{Resonator's dispersive shift $\chi$ as a function of the qubit-resonator detuning $\Delta$ according to LOM and EPR methods, showing the agreement between simulations and the analytical model for $L_\text{j}=10\,$nH.}
    \label{fig:chi_sweep}
\end{figure}

Fig.~\ref{fig:chi_sweep} shows the expected dependence of the resonator dispersive shift with the frequency detuning $\Delta = |\omega_r - \omega_q|$. The comparison between EPR and LOM dispersive shifts at different flux bias conditions, not fully addressed in previous related studies \cite{epr,lom,yuan}, is a useful test that will enable, after a thorough comparison between simulated and real data, to estimate LOM and EPR accuracies at any device working point.
Finally, we quantified through EPR the coupling contributions to the loaded quality factor of the resonator and the qubit, allowing us to estimate their relaxation time through the formula $Q_i = \omega_iT_{1i}$, with $i\in\{r,q\}$ obtaining $T_{1r} = 0.46\,$\textmu s and $T_{1q} = 33.06\,$\textmu s in the zero flux bias condition. The resonator linewidth resulting from this estimate is $k=0.60\,$MHz, which is small enough to make the dispersive shift $2\chi/2\pi$, observable, thus enabling readout at any relevant coupling condition. 

\section{Planar microwave resonator characterization}\label{sec:cpw}
In order to allow for a low-dissipation read-out of qubits, we develop high-$Q$ planar microwave resonators based on aluminium thin-films. For the optimisation of the aluminium film properties and of the microfabrication steps, we designed, microfabricated and characterised a resonator chip consisting of six planar lumped element microwave resonators coupled to a common coplanar waveguide transmission line. The microfabrication of the resonator chip involves two main steps: 1) the aluminium film physical vapour deposition, in which we have deposited a 200$\,$nm thick aluminium film onto a 650$\,$\textmu m thick silicon substrate via magnetron sputtering; 2) the patterning of the aluminium film, in which we have employed mask-based photo-lithography and a wet-chemical aluminium etch to structure the resonators and transmission lines into the aluminium film.

\begin{figure}
    \centering
    \includegraphics[width = 0.4\textwidth]{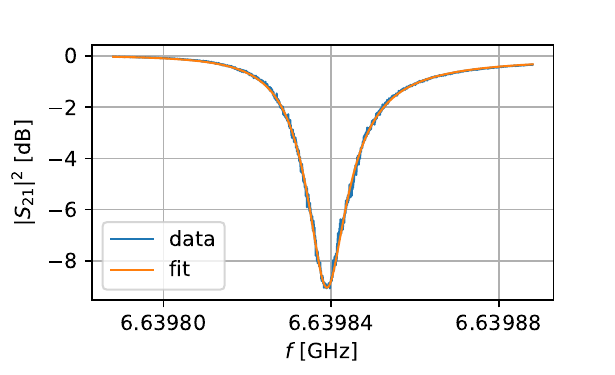}
    \caption{Absolute value of the measured microwave transmission $|S_{21}|^2$ (blue) and of the resonance fit (orange) around the resonance of the second microwave resonator. The read-out power on chip was $P_\mathrm{chip}=-92\,$dBm.}
    \label{fig:resonance_fit}
\end{figure}

\begin{figure}
    \centering
    \includegraphics[width = 0.4\textwidth]{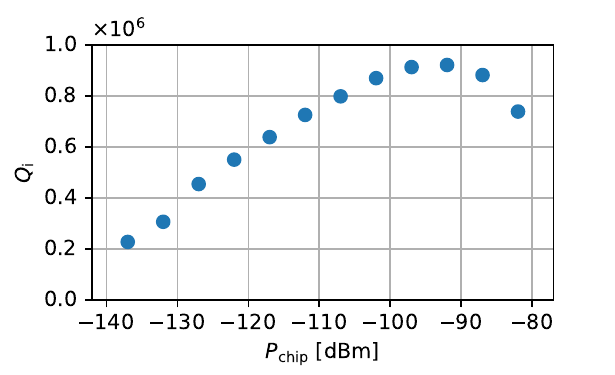}
    \caption{Dependence of the internal quality factor $Q_\mathrm{i}$ on the read-out power on chip $P_\mathrm{chip}$. The shown data corresponds to the second microwave resonator.}
    \label{fig:Qi_power_dependence}
\end{figure}

The final resonator chip has been characterised at a temperature of $30\,$mK by means of a dilution refrigerator. In the microwave transmission spectrum of the chip we have identified six resonances, which we could associate with the chip microwave resonators. From resonance fits to the complex-valued transmission $S_\mathrm{21}$ \cite{Probst2015}, we could extract the resonance frequencies $f_\mathrm{r}$, the internal quality factors $Q_\mathrm{i}$ and the coupling quality factors $Q_\mathrm{c}$ of the resonators. As an example, in Fig.~\ref{fig:resonance_fit}, we show the resonance of the second microwave resonator on chip for an on-chip read-out power of $P_\mathrm{chip}=-92\,$dBm. The read-out power dependence of the internal quality factor $Q_\mathrm{i}$ of the second resonator is depicted in Fig.~\ref{fig:Qi_power_dependence}. We associate the power dependent increase of $Q_\mathrm{i}$ at low powers to the presence of two-level systems \cite{Mueller_2019}, whereas we attribute the decrease of $Q_\mathrm{i}$ at high powers to the breaking of Cooper pairs. At an on-chip read-out power of $P_\mathrm{chip}=-92\,$dBm we find the maximum internal quality factor of $Q_\mathrm{i}=9.2\times10^5$. In order to improve the internal quality factor (in particular in the low power regime), we are currently optimising the fabrication process minimising the presence and the influence of two-level systems.

\section{Conclusions}\label{sec:conclusion}
We finalized the chip layout for the first Qub-IT fabrication and extracted several parameters of interest through EPR and LOM quantization models. The simulations agree with the target values within a few percent and consistency between simulation strategies has been demonstrated. The direct comparison between EPR and LOM dispersive shift evaluation for a flux-tunable transmon provides new insights into assessing their accuracy and prediction agreement.

In preparation for the manufacturing stage,
we optimized the fabrication of coplanar waveguides reaching an internal quality factor well above $10^5$. After chip fabrication, this will be fully characterized and compared to the expected results in order to validate the design and simulation procedure.

\section*{Acknowledgement}
This work was supported by Qub-IT, a project funded by the Italian Institute of Nuclear Physics (INFN) within the Technological and Interdisciplinary Research Commission (CSN5), and PNRR MUR projects PE0000023-NQSTI and CN00000013-ICSC. AG acknowledges support by the Horizon 2020 Marie Sk\l{}odowska-Curie actions (H2020-MSCA-IF GA No.101027746). SP and CB acknowledge support by the University of Salerno - Italy under the projects FRB19PAGAN and FRB20BARON.

\bibliographystyle{IEEEtran}
\bibliography{main}

\end{document}